\documentclass[12pt,preprint]{aastex}
\newcommand{\etal} {et al.\ }
\newcommand{\be}{\begin{equation}}
\newcommand{\ee}{\end{equation}}
\newcommand{\beq}{\begin{eqnarray}}
\newcommand{\eeq}{\end{eqnarray}}

\shorttitle{Simple model for the evolution of coronal loops}
\shortauthors{L\'opez Fuentes \& Klimchuk}

\begin{document}

\title{A simple model for the evolution of multi-stranded coronal loops}


\author{M. C. L\'opez Fuentes\altaffilmark{1,2}, J. A. Klimchuk\altaffilmark{3}}
\altaffiltext{1}{Instituto de Astronom\'{\i}a y F\'{\i}sica del Espacio, CONICET-UBA, CC. 67, Suc. 28, 1428 Buenos Aires, Argentina}
\altaffiltext{2}{Member of the Carrera del Investigador Cient\'{\i}fico y Tecnol\'ogico, Consejo Nacional de Investigaciones Cient\'{\i}ficas y T\'ecnicas (CONICET), Argentina}
\altaffiltext{3}{NASA Goddard Space Flight Center, Code 671, Greenbelt, MD  20771, USA}

\begin{abstract}
We develop and analyze a simple cellular automaton (CA) model that
reproduces the main properties of the evolution of soft X-ray
coronal loops. We are motivated by the observation that these loops
evolve in three distinguishable phases that suggest the development,
maintainance, and decay of a self-organized system. The model is
based on the idea that loops are made of elemental strands that are
heated by the relaxation of magnetic stress in the form of
nanoflares. In this vision, usually called ``the Parker conjecture"
(Parker 1988), the origin of stress is the displacement of the
strand footpoints due to photospheric convective motions. Modeling
the response and evolution of the plasma we obtain synthetic light
curves that have the same characteristic properties (intensity,
fluctuations, and timescales) as the observed cases. We study the
dependence of these properties on the model parameters and find
scaling laws that can be used as observational predictions of the
model. We discuss the implications of our results for the
interpretation of recent loop observations in different wavelengths.

\end{abstract}



\section{Introduction}
\label{intro}


EUV and soft X-ray observations of the solar corona reveal a high
degree of structuring and intermittency, especially in and around
active regions (ARs). Due to the frozen-in condition of the coronal
plasma, the emitting material is ordered by the magnetic field in
elongated features commonly known as coronal loops. The study of the
structure and evolution of loops is therefore fundamental for
understanding coronal dynamics, particularly in relation to the
origin of coronal heating (Mandrini \etal 2000). Historically, the
first attempts to physically describe coronal loops began with what
is known as the Rosner, Tucker \& Vaianna (RTV, 1978) approach,
which considers loops as monolithic structures containing plasma in
static equilibrium. Although soft X-ray loop observations seem to be
consistent with the RTV assumptions (Porter \& Klimchuk 1995), more
recent observations from the Transition Region and Coronal Explorer
(TRACE) suggest that EUV loops are actually too dense to be in
static or quasi-static equilibrium (Aschwanden \etal 2001,
Winebarger \etal 2003). The properties of these loops are more
consistent with the predictions of models based on impulsive heating
(Winebarger \etal 2003, Klimchuk 2006).


TRACE observations also show a high degree of filamentation of the
coronal structure, suggesting that what appear to be monolithic
loops can actually be collections of unresolved individual strands
with more or less independent evolutions. Although the question of
whether coronal loops are isothermal or multithermal has been a
matter of a heated debate during several years (Aschwanden \&
Nightingale 2005, Schmelz \& Martens 2006), recent observations seem
to indicate that both kinds of loops actually occur (Schmelz \etal
2009). Combining all the available information, there are strong
arguments in favor of many loops being made of unresolved strands
that evolve more dynamically than what quasi-static models predict
(for a recent review on the subject see Klimchuk 2009).


There are two main families of mechanisms proposed to explain the
origin of coronal heating: the so called AC and DC heatings.
Generally speaking, AC refers to the dissipation of MHD waves, while
DC relates to the dissipation of highly stressed magnetic fields.
The latter involve the release of free magnetic energy at locations
of strong gradients in the magnetic field, i.e., current sheets.
Ultimately, the source of energy is the distortion of the magnetic
structures by convective motions. This is precisely the mechanism
envisioned by Parker (1988). He proposed that, since the footpoints
of magnetic strands are being displaced by photospheric granular
motions, there must be a continuous increase of the magnetic stress
between neighbor strands. These stresses accumulate until some
threshold is reached and reconnection occurs. In the process, stored
magnetic energy is released and the plasma is heated. Because of
similarities with the usual flaring process and the energy scale
involved, these reconnection events are usually called nanoflares
(Cargill 1994, Cargill \& Klimchuk 2004). One key point of the
process described above is the existence of a threshold or critical
value for the reconnection to occur. A theoretical support for the
existence of this threshold is given in Dahlburg \etal (2005, 2009).
They found that when certain field misalignment is reached the
secondary instability provides the mechanism for a rapid energy
release.


The intermittent and filamentary behavior, and the possible presence
of power law distributions (see e.g., Aschwanden \& Parnell 2002),
led some authors to propose that the solar corona might be a good
example of a self-organized critical (SOC) system (Bak 1990).
Beginning with the seminal work of Lu \& Hamilton (1991), there have
been an important number of studies involving the use of SOC models
in relation to the coronal heating problem (for a review on SOC
models see Charbonneau \etal 2001).


In our recent paper, L\'opez Fuentes \etal (2007), we studied the
evolution of coronal loops observed with the Solar X-ray Imager
(SXI) on board the Geosynchronous Operational Environmental
Satellite 12 (GOES-12). Since this instrument observes the sun
almost continuously, it allowed us to follow the evolution of a set
of coronal loops from birth to disappearance. We found that the loop
evolution can be separated into three main phases, which we called:
\textit{rise}, \textit{main} and \textit{decay}. As we indicated
there and began to explore in Klimchuk, L\'opez Fuentes \& DeVore
(2006), the observed evolution can be related to the development,
maintenance and destruction of a critical system. Here, we develop a
simple model based on a cellular automaton (CA) that reproduces the
observed evolution. We model the plasma response to the heating
using the EBTEL hydrodynamics code (Klimchuk \etal 2008). We then
create synthetic light curves taking into account the
temperature-dependent response of the SXI instrument.  We compare
these with the observed light curves and analyze how the different
parameters of the model influence the properties of the obtained
light curves, such as the characteristic intensities and timescales,
which we studied in detail in L\'opez Fuentes \etal (2007) for the
observed cases.


In Section~\ref{obs_loops} we sumarize the results from L\'opez Fuentes \etal (2007) and show examples of observed light curves that will be later compared with the model. In Section~\ref{model} we describe the CA model and in Section~\ref{results} we analyze our results. We discuss and conclude in Section~\ref{discussion}.


\section{Observed loop evolution}
\label{obs_loops}


In L\'opez Fuentes \etal (2007) we studied the evolution of a set of
coronal loops observed with the Solar X-ray Imager (SXI) on board
GOES-12 (Hill \etal 2005). This instrument has the advantage of
observing the sun almost continuously, allowing us to follow the
whole evolution of loops. Previous similar studies based on
Yohkoh/SXT or TRACE observations were limited by the occultation
times of the satellites and times of interrupted observation due to
high radiation levels (e.g., when passing through the South Atlantic
anomaly), therefore reducing the continuous observation times to a
small portion of the loop lifetime (see e.g. Porter \& Klimchuk
1995).


We analyzed a set of 457 images taken with the telescope in the
filter position ``open'', which is most sensitive to emission from
plasmas below 2MK. The analyzed data span three days (August 26-28,
2003) of observations with a cadence of approximately 6 images per
hour. From these data we selected and tracked the evolution of 7
coronal loops. Following a procedure that is thoroughly described in
L\'opez Fuentes \etal (2007), we obtained light curves of the
intrinsic (background subtracted) intensity of each of these loops.
We found that the loop evolution can be separated into three parts:
a \textit{rise} phase during which the loop intensity slowly
increases, a \textit{main} phase during which the intensity remains
approximately constant (or its long term variation is much less
pronounced), and a \textit{decay} phase during which the intensity
decreases until the loop is almost indistinguishable from the
background. In Figure~\ref{lightcurves} (upper panels) we reproduce
adapted versions of two of the light curves studied in
L\'opez Fuentes \etal (2007). In the original analysis we applied a
procedure to remove the background contribution that left a residual
component of long term variation (see discussion in Section 3.1 in
L\'opez Fuentes \etal 2007). For an easier comparison with the model
light curves, here we removed the residual background by subtracting
in each case a linear function of time, so that the initial and final
intensities of the loops are reduced nearly to zero.


We found that the durations and characteristic timescales of the
rise and decay phases are longer than the expected cooling times. We
concluded that this is consistent with two alternate scenarios. In
the first one, loops are monolithic and the heating source varies
very slowly with time, so the plasma responds quasi-statically. In
this way, the three evolution phases would reflect the phases of the
long term change of a slowly varying heating source. In the second
scenario, loops are made of unresolved and independent magnetic
strands which are heated by impulsive events such as nanoflares.
Since the observed loop emission is the summed contribution of the
constituent strands, the different phases would be directly related
to the variation of the number and intensity of the energy
dissipation events that heat the strands. We argued that under this
particular scenario the observed loop evolution can be related to
the development, maintenance and destruction of a self-organized
system. In the next sections we present and analyze a model that
reproduces precisely this kind of behavior.


\section{Cellular automaton model}
\label{model}

\subsection{General description}
\label{description}

In this Section we describe the construction of a cellular automaton (CA) model that reproduces the main features of the observed loop evolution. Motivated by the arguments given in Section~\ref{intro}, the model simulates the process described there regarding the scheme: motion of magnetic strand footpoints $\to$ increase of magnetic stress $\to$ energy release by reconnection. Figure 2 shows a cartoon that illustrates the main idea.

Let us begin with a bunch of thin parallel magnetic flux tubes or
strands that connect two different regions of the solar surface
(panel (a)). For simplicity the strands are drawn straight, so the
top and bottom planes correspond to two relatively distant
photospheric regions of opposite magnetic polarity. As time goes on,
strand footpoints are horizontally dragged by random photospheric
motions, so the strands become progressively more misaligned. This
produces growing magnetic stresses and intensifying current sheets
in the region separating adjacent strands (panel (b)). For two given
neighbor strands, the misalignment angle increases until a critical
value is reached (i.e., the secondary instability develops) and the
energy is released in the form of an impulsive event. Consequently,
the field between the two strands relaxes.  This sudden change in
the field may occasionally cause new instability conditions between
these strands and their other neighbors. Under some conditions
instabilities propagate farther producing avalanches of events
involving many strands. This is the typical situation considered in
SOC models.  For clarity, we adopt the following terminology.  The
basic interaction between two strands that occurs each time the
critical value is reached is called a ``reconnection event." The
total release of energy into a single strand by its reconnection
with one or more neighbors over a short period of time is called a
``nanoflare." This follows the convention used by the loop modeling
community.

The system continues to evolve in response to the footpoint driving
until there is a rather abrupt appearance
of a critical state in which the energy input by the photospheric
motions is statistically balanced by the dissipation due to
reconnection events. In the critical state the total energy of the
system fluctuates intermittently around a constant level. Local
intermittency results from the occurrence of nanoflares of different
sizes.

The idea is then to develop a numerical algorithm that
quantitatively reproduces the above scheme. First of all, we need to
translate footpoint displacements to magnetic field stresses. To
accomplish this we begin with the simple scenario shown in Figure 2
(c). Let us concentrate on one particular strand and, for
simplicity, suppose that only one footpoint (the lower one) is being
displaced. If the particular strand has a vertical magnetic field
strength $B_{v}$, any horizontal displacement of its footpoint will
produce a new horizontal component of the field

\be
\delta B_{h}=B_{v}\tan\theta,
\ee

\noindent where $\theta$ is the angle that the now inclined strand forms
with the vertical direction. It is easy to see that if the strand has a
length $L$ and the footpoint has been displaced a distance $d$, then
$\tan\theta = d/L$. Therefore:

\be \label{driver_eq} \delta B_{h} = B_{v}d/L  \label{eq:dBh}. \ee

\noindent The distance $d$ is a characteristic length scale of the
photospheric motions, which can be associated with the sustained
displacement of a strand footpoint during the turnover time of a
convective granule ($\approx  10^{3}$ s). Assuming a typical
photospheric velocity $v \approx 1$ km/sec, results in
$d \approx 1000$ km. After a number of successive displacements
the strand would acquire a horizontal field

\be B_{h} = \Sigma \delta B_{h}. \ee

\noindent At this point we have assumed that $B_h$ increases with
each and every step. This is only true if the new step is in the
same direction as the previous step or if the change in direction
causes the strand to wrap around a neighbor.  The latter will
generally be true if the random walk step size is larger than the
characteristic separation of neighbor footpoints.  However, even
then, some steps will cause the strand to back track and $B_h$ will
decrease. We treat this more realistic situation below.

As a consequence of the random walk, the horizontal field $B_{h}$
will gradually increase until the critical condition referred to
above is reached. If we call $\theta_{c}$ the critical angle of
misalignment between two adjacent strands, it is easy to see that
there is an associated critical value of the difference in their
horizontal field components:

\be
B_{c}=B_{v}\tan\theta_{c}.
\ee

\noindent Note that $\theta_{c}$ refers to the relative angle
between the adjacent strands and not the inclination from vertical,
$\theta$, in Equation 1.

Assuming 1D displacements (see Section~\ref{implementation}), we
relate the misalignment between two neighbor strands (with indices
$p$ and $q$) to the difference of their horizontal field components
$\Delta B_{h}=B_{h}(p)-B_{h}(q)$. The fulfillment of the critical
condition implies $|\Delta B_{h}|>B_{c}$. If we assume that all the
field is equally distributed after reconnection, the new (primed)
horizontal field in each strand will respectively be:

\be
\label{field_distr_eq}
B'_{h}(p)=B_{h}(p)-\Delta B_{h}/2~~~\textrm{and}~~~B'_{h}(q)=B_{h}(q)+\Delta B_{h}/2.
\ee

The released energy corresponds then to the difference in the total magnetic energy before and after reconnection. It can be easily demonstrated that

\be \label{release_eq} \Delta E=(E'-E)=\left( \frac{1}{8\pi} \right)
\frac{(\Delta B_{h})^{2}}{4} \ee

\noindent per unit volume, where we have assumed that the energy is
divided equally between the plasma in the two strands.

\subsection{Model implementation}
\label{implementation}


For the numerical implementation of the model we use a 1D array of
$N_{m}$ points. Each point corresponds to a single magnetic strand
and, initially, all strands have horizontal magnetic field
$B_{h}=0$. At each time step, a small random amount of field $\delta
B_{h}$ is added to each site according to a statistical distribution
that simulates the photospheric displacements of strand footpoints.
Although the field increase in the model is 1D, we apply a
distribution based on 2D footpoint displacements. The idea is
illustrated in Figure~\ref{distribution}-a. If one compares the
position of a strand footpoint at time step $(t-1)$ with its
position at time step $t$ (gray arrow), the probability that during
the following time step, $(t+1)$ (black arrows) the footpoint moves
farther away increasing the horizontal magnetic strength is
approximately 3/4, while the probability that it moves back on its
previous track, canceling (or almost canceling) the previous
displacement is roughly 1/4. Therefore, we add the $\delta B_{h}$
values according to the distribution shown in
Figure~\ref{distribution}-b. Values from the gaussian distribution
centered on $d_{0}$ are 3 times more probable than values given by
the gaussian centered on $-d_{0}$. The distance $d_{0}=1000$ km is
the average horizontal displacement of a strand footpoint during the
turnover time of a granular cell (in our scheme, the duration of
each time step). The width of the gaussians is 20\% of $d_{0}$. As
explained in the previous Subsection, the random displacements taken
from this distribution translate into magnetic inputs through
Equation~\ref{driver_eq}.


It is easy to see that under this simple 1D scheme all strands will
eventually acquire positive values of $B_{h}$. To keep the
simplicity of the model and to be consistent with the fact that
neighbor strand footpoints tend to move in different directions, we
alternate the sign of the displacements $d$ among neighbor strands.
This means that, identifying each mesh site with an index $i$, we
add the random values obtained from the distribution to strands with
$even~i$, and we subtract them from strands with $odd~i$. Thus, in
the long run $even$ strands acquire positive $B_{h}$ while $odd$
strands acquire negative $B_{h}$. This alternation of $\delta B_{h}$
simulates in 1D the magnetic stress produced by footpoints moving in
different random directions.


The efficiency of the process that injects ``horizontal component''
of the magnetic field through the motion of the footpoints depends
on the average distance between neighboring strands. If the strands
are closer than $d_{0}$, then the process can be considered
efficient in tangling the strands and accumulating magnetic stress.
However, if there is no injection of new flux, then the distance
between strands will increase as the footpoints disperse.  The
driving process will become progressively less efficient. To
simulate the decay of the driving mechanism due to footpoint
dispersion, at a chosen point during the system evolution we make
the distribution shown in Figure~\ref{distribution}-b slowly evolve
to the symmetrical distribution shown in
Figure~\ref{distribution}-c. In the final state, the driver
distribution is completely inefficient at creating new $B_{h}$. As
we will see below, this evolution of the driver leads to the decay
of the system. The start time and the duration of the evolution from
the initial to the final distribution are set as parameters of the
model.


In Subsection~\ref{description} we defined
$B_{c}=B_{v}\tan\theta_{c}$. Therefore, $\tan\theta_{c}$ is one of
the relevant parameters of the model. At each time step, after
adding the random $\delta B_{h}$ values to all the strands, we
compare $B_{h}$ on each strand with its immediate neighbors. For a
strand having index $i$ we compute the difference $\Delta B_{h}(i) =
B_{h}(i)-B_{h}(i+1)$.  We do this for all the strands in the array.
We use periodic boundary conditions, so strand $i=N_{m}$ is compared
with strand $i=1$, and in that case $\Delta B_{h}(N_{m}) =
B_{h}(N_{m})-B_{h}(1)$. Next, we identify all the indices where
$|\Delta B_{h}(i)| > B_{c}$. For all the sites where this happens,
we redistribute the horizontal field according to
Equation~\ref{field_distr_eq}.

As we discussed in Subsection~\ref{description}, each field
redistribution is associated with a reconnection event that frees an
energy $\Delta E$, in erg cm$^{-3}$, according to
Equation~\ref{release_eq}. This energy is used to heat the plasma in
both strands involved in the event. Before proceeding to the next
time step we repeat the previous test and corresponding field
redistribution as many times as necessary until all $|\Delta
B_{h}(i)|$ values are less than $B_{c}$. Once this state is reached,
the system is allowed to evolve to the next time step, a new set of
random $\delta B_{h}$ values is added to the $B_{h}$ of the strands,
and the process described above is repeated.

During the evolution of the system we keep record of the energy
dissipated in each strand at each time step. Here, we assume that
the reconnection events ($\Delta E$) occur on a timescale that is
much shorter than the time step duration ($\approx 10^3$ s). As we
defined in Section~\ref{description}, the sum of all the $\Delta E$
``energy packages'' that heat the plasma in the strand during a
given time step is called a nanoflare. The final output of the CA
model is the array of nanoflare energies dissipated in each strand
as a function of time.


In the above description we assumed that when the critical condition
($|\Delta B_{h}| > B_{c}$) is fulfilled all the available field
$\Delta B_{h}$ is distributed (reconnected) among the two strands.
This is not realistic.  On the Sun, each strand of the tangled field
is wrapped around many other strands.  It is only the horizontal
field component along the section of mutual contact that is free to
reconnect and heat the plasma.  We account for this in our simple
model by introducing a parameter $f$.  The initial response to the
reconnection is to change the horizontal field locally by an amount
$\Delta B_{h}/2$.  This change is subsequently spread out along the
entire length of strand. The adjustment happens at the Alfven speed
and is very rapid. Assuming a minimum Alfven speed of
$1000~km~s^{-1}$ for the solar corona gives a time of $100~s$ or
less. The end result is that the horizontal field changes by
$f\Delta B_{h}/2$ along the entire strand. It can be demonstrated
that in this case the energy release on each of the two strands per
unit volume is

\be
\Delta E = \left(\frac{1}{8\pi}\right)f\left(1- \frac{f}{2} \right) \frac{(\Delta B_{h})^{2}}{2}.
\label{release_eq2}
\ee

Note that the parameter $f$ affects the time required for a pair of
adjacent strands to once again reach the critical condition through
continued driving. If $f$ is small, the strands will never be far
from critical and nanoflares will recur frequently. If $f$ is large,
it will take many steps to once again reach the critical level of
stress. Of course the nanoflares will be weaker in the former case
than in the latter, so the temporally averaged heating will not be
greatly different. There are small differences though, reflected in
the $f$ dependence of equation (\ref{energy_eq}) discussed later.


For the analysis, we vary the values of the model parameters
$B_{v}$, $L$, $\tan \theta_{c}$, $f$, $N_{m}$, and $\tau$.  The
parameter $\tau$ is the duration of the nanoflares as defined later
in Section~\ref{plasma}. 
Table~\ref{table1} shows the different values used. For the vertical
magnetic field we use values in the range 20-120 Gauss, typical of
AR loops (Mandrini \etal 2000). We vary the length of the loops also
considering typical AR values (20-120 Mm). The tangent of the
critical angle ($\tan \theta_{c}$) varies between 0.1 and 1.0,
corresponding approximately to angles: 6 deg, 11 deg, 22 deg, 31
deg, 39 deg, and 45 deg. The reconnection factor $f$ varies between
0.1 and 1.0, meaning that 10\% to 100\% of the available magnetic
stress is released by reconnection. The total number of strands,
$N_{m}$, goes from 20 to 1000, which is roughly the number of
elemental magnetic flux tubes believed to be present in a coronal
loop.  We vary one parameter at a time keeping the rest of them at
the intermediate values shown in bold face in Table~\ref{table1}.


As we stated above, the output of the CA model is an array of nanoflare
energies dissipated in each strand as a function of time, with time
steps of $\approx 10^3$ s. Figure~\ref{example} helps to illustrate
the typical evolution of the system. For the example we use the
parameter combination shown in bold face in Table~\ref{table1} and a
total duration of 72 timesteps ($\approx 20~hr$). The upper panel
shows the evolution of the mean energy of the system, defined as the
sum of the magnetic energies of all the strands ($B_{i}^{2}/8\pi$)
divided by the number of strands. The middle panel shows the
evolution of the mean nanoflare energy, which is the sum of the
energies of all nanoflares occurred at each timestep divided by the
number of strands. The evolution is such that the number of
reconnection events increases from zero (rise phase) to a
quasi-stationary level (main phase). In the example of
Figure~\ref{example} this transition occurs around time step number
20. The system reaches this state abruptly, not asymptotically, and
maintains it for as long as the driver follows the distribution of
Figure~\ref{distribution}-b. Notice that once the system reaches the
main phase the magnetic energy (Figure~\ref{example}, upper panel)
maintains an approximately constant level, except for fluctuations
due to the nanoflares. As soon as the driver begins to evolve
towards the distribution of Figure~\ref{distribution}-c, the number
of events decays until they almost disappear at the end of the
system's evolution (Figure~\ref{example}, middle panel). Likewise,
once the decay phase begins the fluctuations of the system energy
begin to decrease rapidly and stop. Summarizing, the evolution of
the full system qualitatively follows the evolution of the observed
coronal loops in terms of $rise$, $main$, and $decay$ phases.

\subsection{Plasma response and light curve construction}
\label{plasma}


Since our main motivation is to compare the output of the CA model
with the loops studied in L\'opez Fuentes et al. (2007), the next
step is to transform this output into an observed signal such as the
light curve of a coronal loop. To do this, we simulate the plasma
response using the EBTEL (Enthalpy-Based Thermal Evolution of Loops)
model, developed by Klimchuk, Patsourakos \& Cargill (2008). The
temperature and density of the coronal plasma tend to be reasonably
uniform along the magnetic field, and EBTEL computes the evolution
of the spatially-averaged values.  This approximate 0D solution is
quite similar to the exact solution given by 1D hydrodynamic codes,
but it takes 3-4 orders of magnitude less time to compute. EBTEL is
therefore ideally suited for our study in which we simulate the
evolution of thousands of individual strands.

The main input of EBTEL is the heating rate profile, i.e., the
variation of the heating rate with time in the strand.  Here, we
assume that nanoflares have a triangular profile of a given duration
$\tau$. The upper panel of Figure~\ref{ebtel} shows a series of
``triangular'' nanoflares that heat one of the strands of the CA
model during a portion of the main phase. The time integral of
each triangular heat function corresponds to the total energy
provided by the particular nanoflare. The duration of the nanoflares shown
in Figure~\ref{ebtel} is 1000 s. This means that, in this case, the
available heat is distributed (with a triangular profile) along a
full time step of the CA model. The other two panels of
Figure~\ref{ebtel} show the output of EBTEL: the plasma temperature
and density. During the time interval covered in the panels there
are both isolated events (like the nanoflare starting at
$t=4.3\times10^{4}$ s, gray arrow), and ``trains'' of events occurring in rapid
succession (such as the one starting at $t =4.7\times10^{4}$ s, black arrow).
Notice that while the plasma relaxes almost fully after the isolated event,
during the succession of multiple events the plasma is still in a state
of high density and temperature due to the previous nanoflare when the
next event begins. As we will see in the following sections, this has
consequences in the way the plasma emits and the observed loop
intensity evolves.

The loops studied in L\'opez Fuentes \etal (2007) have measured
densities in the range 0.9-1.9$\times10^{9}$ cm$^{-3}$, and measured
temperatures between 1.2 and 2.1 MK. These values are reproduced by
parameter values in the ranges given in Table~\ref{table1}.


Once we have the temperature and density evolution of each strand,
we compute the observed intensity using the emission measure and the
known temperature dependent sensitivity of the instrument. The
emission measure per pixel of a single unresolved strand is
$EM=n^{2} \Delta V$, where $\Delta V$ is the volume of the strand
portion covered by a pixel. This is roughly $\Delta V=
l_{pix}A_{s}$, where $l_{pix}$ is the pixel dimension and $A_{s}$ is
the cross sectional area of the strand.  The contribution to the
intensity in a pixel is then $I = E\!M~S(T)$, where $S(T)$ is the
SXI response function.  We sum the contributions of all the strands
to determine the intensity of the loop and, finally, we multiply
this intensity by a factor that accounts for the proportion of the
loop covered by a single pixel. The final output is the observed
intensity of the loop per pixel per second as a function of time.
The data are sampled using the instrument temporal resolution to
simulate observed light curves. In Figure~\ref{example}, lower
panel, we show a light curve obtained in this way for the model
example of the upper and middle panels. Comparing the lower and
middle panels it can be noticed that the light curve follows (though
smoothed) the general fluctuations produced by the summed
contribution of the nanoflares.


\section{Results}
\label{results}

\begin{figure*}
\centering
\hspace{0.cm}
\includegraphics[bb=70 240 570 610,width=17cm]{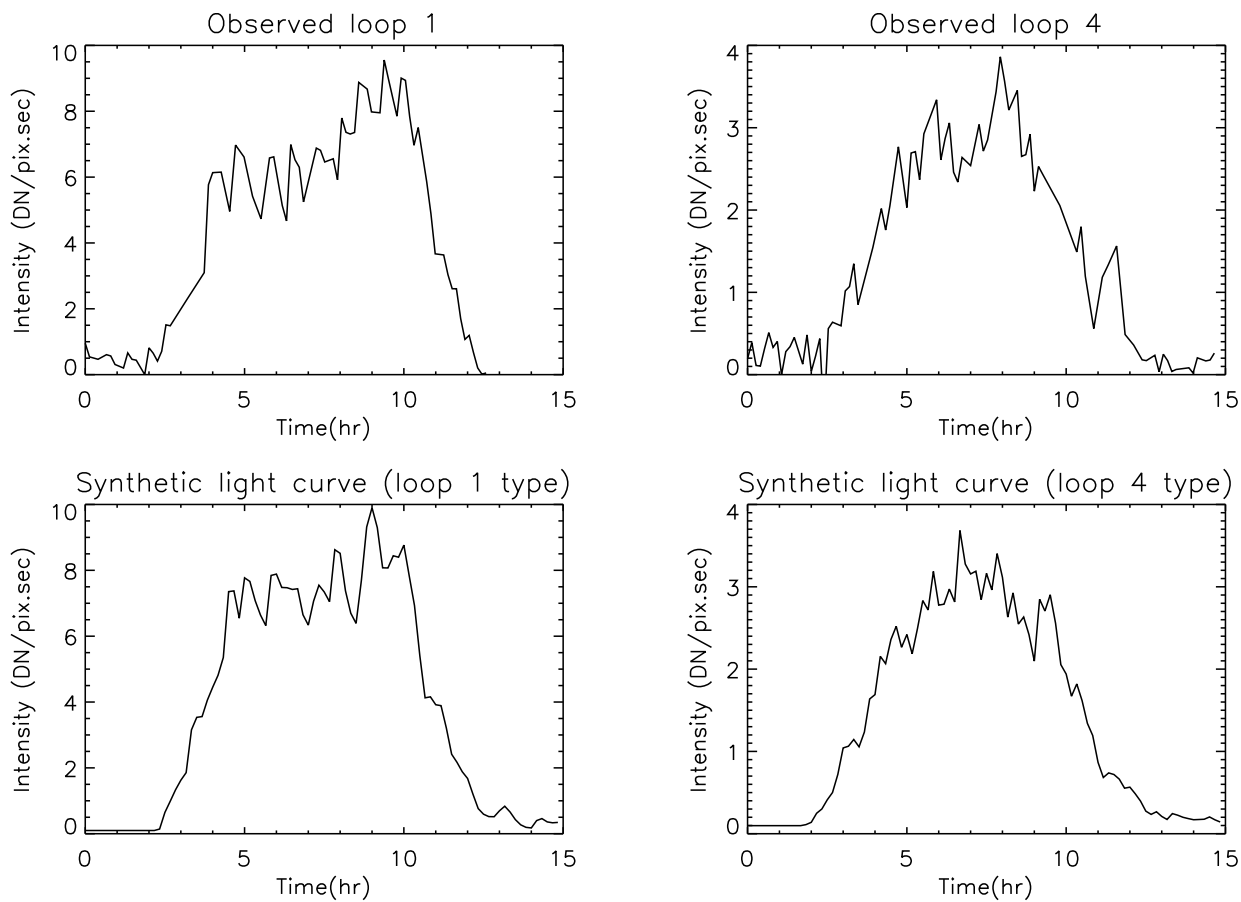}
       \caption{Evolution of observed and synthetic X-ray loops. The light curves obtained with the CA model (lower panels) reproduce the main properties of the evolution of the observed loops (upper panels). See Section~\ref{results}.}
         \label{lightcurves}
\end{figure*}

The lower panels of Figure~\ref{lightcurves} show two model light curves obtained in the way described in Section~\ref{plasma}. The cases shown have the same properties as the observed light curves of the upper panels. Given the random nature of the nanoflares, we do not expect an exact fit of the model to the observed data. For the comparison we instead consider the light-curve properties studied in L\'opez Fuentes et al. (2007) (see Table~\ref{table2}).

The values of the model parameters used to construct the synthetic
light curves of Figure~\ref{lightcurves} are, for Loop 1:
$B_{v}=37$~G, $\tan \theta_{c}=0.15$ and $f=0.2$; and for Loop 4:
$B_{v}=20$~G, $\tan \theta_{c}=0.2$ and $f=0.1$. For both cases we use
$N_{m}=100$, and $\tau=1000$~s. For the loop length and diameter we use
the measures obtained in L\'opez Fuentes et al. (2007): $L=110$ Mm for
loop 1, $L=70$ Mm for loop 4, and a diameter of $1.8 \times 10^9$ cm for
both.

We compute for both observed and synthetic light curves the mean intensity 
of the main phase ($I_{m}$) and the duration and timescales of the rise 
and decay phases. To obtain these values we first identify, by visual 
inspection, the approximate start and end times of each phase. Using the 
same definitions as in L\'opez Fuentes \etal (2007), we have for the rise
phase timescale:

\be
T_{rise} = \frac {\bar{I}}{(\Delta I/\Delta t_{rise})}.
\label{trise_eq1}
\ee

\noindent Here, $\bar{I}$ is the mean intensity during the rise
phase, $\Delta I$ is the intensity variation and $\Delta t_{rise}$ is the
duration. A similar definition is used for the decay phase. As a
proxy for the amplitude of the intensity fluctuations, we compute
the root-mean-square ($RMS$) of the intensity variation during the
main phase relative to its ten-point running average. In
Table~\ref{table2} we compare these properties for observed and
modeled light curves. It can be seen that the model reproduces the
light curve properties consistently.

We remind the reader that the observed light curves shown in
Figure~\ref{lightcurves} have been further processed (residual
background removed, see Section~\ref{obs_loops}) from the versions
studied in L\'opez Fuentes \etal (2007). Therefore, the properties
listed in Table~\ref{table2} do not necessarily coincide with the
corresponding values presented in the previous paper.

Regarding the durations and timescales it is interesting to see that,
since the light curves have initial intensities very close to zero,
$\bar{I}$ in the Equation~\ref{trise_eq1} is approximately $\Delta I/2$.
Therefore,

\be
\label{trise_eq2}
T_{rise} \approx \frac{\Delta t_{rise}}{2}.
\ee

\noindent We then expect the timescales to be approximately one half
of the durations (compare the corresponding values in
Table~\ref{table2} for both the rise and decay phases). The relation
from Equation~\ref{trise_eq2} is not valid, in general, for the
light curves studied in L\'opez Fuentes \etal (2007), in which there
is still a remnant background component of the intensity.


To investigate how the model parameters determine the properties of
the loops, we vary each of them separately according to
Table~\ref{table1} while keeping the rest fixed at the values shown
in bold face (see Section~\ref{model}). In Figure~\ref{energy} we
show log-log plots of the the mean energy per unit volume released in reconnection
events, $\Delta E$ (see Equation~\ref{release_eq2}),
versus different model parameters. The lines are least-square fits
of the data, and their slopes are indicated in the panels.
According to these plots, in our CA model the energy of the reconnection
events approximately scales with $B_{v}^{2.01}$ and $(\tan \theta_{c})^{1.89}$.
This can be easily explained by noting that in Equation~\ref{release_eq2}, when
reconnection occurs, $\Delta B_{h} \approx B_{c} = B_{v}\tan \theta_{c}$.
Therefore, it is expected that $\Delta E \propto B_{v}^2~(\tan \theta_c)^2$.

The horizontal field decreases by an amount $f\Delta B_{h}/2$ during
a reconnection event.  Several timesteps are typically required to
rebuild the field to the critical level whereupon another event occurs.
From equation (\ref{eq:dBh}) we see that the number of recovery timesteps
is given by

\be N_{r} \approx \frac{f}{2} \frac{L}{d}~\tan \theta_c.
\ee

\noindent From this we obtain the mean nanoflare heating rate
(averaged both in time and along the strand):

\be \left< Q \right> \approx \frac{\Delta E}{N_r t_{s}} = \left(1 -
\frac{f}{2} \right) \frac{v B_{v}^{2} \tan \theta_{c}}{L} ,
\label{energy_eq} \ee

\noindent where $t_{s}$ is the time step duration and $v = d /
t_{s}$ is the photospheric driving velocity. We have assumed that
there is one reconnection event per nanoflare, whereas often there
are two, but this is not important for identifying scaling
relationships.  In Figure~\ref{nanoflares} we show log-log plots of
the mean heating rate versus $B_{v}$, $\tan \theta_c$, and $L$. The
dependence on the parameters is: $B_{v}^{2.02}$, $L^{-1.1}$, and
$(\tan \theta_{c})^{0.97}$, which is clearly explained by
Equation~\ref{energy_eq}. It is easy to see that $\left< Q \right>$
does not depend on the number of strands, $N_{m}$, since both the
driver and the critical condition are applied to the strands
individually.


We can also analyze scaling relations for the properties of the
light curves obtained with the model. For each combination of the
parameters from Table~\ref{table1} we obtain synthetic light curves
following the procedure described in Section~\ref{plasma}. In
Figure~\ref{main} we present log-log plots of the mean intensity of
the main phase, $I_{m}$, as a function of the same parameters as in
Figure~\ref{nanoflares}. The scalings found in this case are:
$B_{v}^{2.37}$, $L^{-0.8}$, and $(\tan \theta_{c})^{1.15}$. Clearly,
$I_{m}$ is related to the nanoflare energy, but the dependence is
complex and is affected by both the detailed response of the plasma
and the nonuniform temperature sensitivity of the instrument.
$I_{m}$ is not linearly proportional to the total radiation emitted
by the strand. To understand the origin of the dependence, let us
now consider the case of quasi-static equilibrium.  This is a
reasonable representation if the time interval between nanoflares is
short compared to a cooling time.  When such conditions apply, the
conductive and the radiation terms of the energy balance equation
are comparable in magnitude to each other and to the mean heating
rate, $\left< Q \right>$ (Vesecky \etal 1979):

\be \left< Q \right> \approx \frac{2}{7} \kappa_{0}
\frac{T^{7/2}}{L^{2}} \approx n^{2} \Lambda_0 T^{b}.
\label{balance_eq} \ee

\noindent Here, $\kappa_0$ is the coefficient of thermal
conductivity, $n$ is the electron number density, and $\Lambda_0$
and $b$ are constants. From the first part of
Equation~\ref{balance_eq} we can obtain for the temperature:

\be T \propto \left< Q \right>^{2/7}L^{4/7} \propto B_{v}^{4/7}
L^{2/7} (\tan \theta_c)^{2/7} . \label{temperature_eq} \ee

\noindent For the temperature range of the SXI loops studied in
L\'opez Fuentes \etal (2007), a reasonable single power value for
the radiation loss is $b \approx -0.5$ (RTV, 1978). Using this, and
equations~\ref{balance_eq} and~\ref{temperature_eq} we have:

\be
n \propto \frac{T^{2}}{L} \propto B_{v}^{8/7} L^{-3/7} (\tan \theta_c)^{4/7}.
\label{density_eq}
\ee

The observed loop intensity is $I = n^2 S(T)$, where $S(T)$ is the
response function of the instrument, which we can express as $S(T) =
S_{0} T^{a}$. We found that $a \approx 0.7$ for the temperature
range of the loops studied in L\'opez Fuentes \etal (2007).
Replacing equations~\ref{temperature_eq} and~\ref{density_eq} in the
expression for the loop intensity, we obtain finally:

\be
I_m \propto B_{v}^{2.69} L^{-0.66} (\tan \theta_c)^{1.34}.
\label{intensity_eq}
\ee

\noindent This expression is consistent with the scalings in
Figure~\ref{main}.  Differences are to be expected since the strands
are sometimes far from quasi-static equilibrium (Fig. \ref{ebtel}).


It might be expected that the intensity of the main phase depends on
$N_{m}$, but this is not the case. The reason is as follows. The
intensity contribution of each strand is proportional to the volume
of the strand subtended by the instrument pixel, which is
proportional to the strand cross-section. Since we compute the
strand cross section as the observed cross-section of the loop ($1.8
\times 10^{9}$ cm) divided by $N_{m}$, an increase of $N_{m}$ means
a proportional decrease of each strand intensity, so the summed
intensity of all strands remains unchanged.


As we mentioned above, another important property of the light
curves is the amplitude of the intensity fluctuations, which we
quantify from the $RMS$ of the intensity variation during the main
phase. These fluctuations are related to the frequency and relative
sizes of the nanoflares. As we briefly discussed in
Section~\ref{model} the factor $f$, which relates to the fraction
of the strand length that is involved in the reconnection,
determines the waiting time for two neighbor strands to recover the
critical condition. If $f$ is small the waiting time will be short
and the nanoflares will tend to be smaller and more frequent. On the
other hand, as $f$ tends to 1 the nanoflares will be larger and less
frequent. Therefore, it is expected that the intensity fluctuations
increase with $f$. The upper panel of Figure~\ref{rms} shows
precisely that.

The intensity fluctuations are also affected by the number of
strands, $N_{m}$. Since the intensity of the loop is the summed
contribution of all the strands, given the random nature of the
release events, the larger the number of strands, the smoother the
signal. This is shown in Figure~\ref{rms}, middle panel. We see that
the $RMS$ intensity fluctuation decreases as the number of strands
increases. The effect levels off once a sufficiently large number of
strands are present.

We also explore how the nanoflare duration, $\tau$, affects the
intensity fluctuations. The different values used are presented in
Table~\ref{table1}, last row. The results are shown in
Figure~\ref{rms}, lower panel, where it can be seen that the $RMS$
fluctuation also decreases as $\tau$ increases. The reason is that
individual strands evolve more slowly as the nanoflares get longer.
For long enough nanoflares the strands behave in a quasi-static
manner.  As is the case with $N_{m}$, the effect levels off once the
nanoflares are sufficiently long.

Finally, although the dependence is much weaker than for
the previous parameters, the $RMS$ variation tends to increase with
all the parameter changes that increase the intensity, i.e., the
increase of $B_{v}$ and $\tan \theta_{c}$, and the decrease of $L$.
This can be understood in terms of the time it takes the plasma to
cool fully after a nanoflare.  The cooling time varies inversely
with the nanoflare energy and directly with the loop length (Cargill
\& Klimchuk 2004).  Shorter cooling times result in more bursty
emission and greater fluctuation amplitudes.  The increase of the
fluctuation amplitude with the increasing intensity is consistent
with the observational results obtained recently by Sakamoto \etal
(2008, and references therein) and discussed also in Vekstein
(2009). They also found that the amplitude of the intensity
fluctuations of Yohkoh/SXT loops is approximately 8\% of the value
of the loop intensity. We find similar values in our own
observations and the results obtained with the CA model (see the
main phase $RMS$ variation in Table~\ref{table2}).


Another important property of the light curves is the rise duration,
$\Delta t_{rise}$, which is the time it takes to reach the main
phase after the intensity first starts to rise.  Note that the
intensity does not increase immediately when footpoint driving
begins (see Figure~\ref{example}, middle and lower panels), because
it takes time for the strands to reach the critical condition. The
first strands to reach it are those that by chance have
larger-than-average step sizes. We define the waiting time $T_w$ to
be the elapsed time between the start of the evolution ($t=0$) and
the main phase beginning. It can be easily seen that $T_{w}$ is
simply the number of steps required to reach the critical condition,
$|B_{h}| = B_{c}/2$, where the factor of 2 is because odd and even
strands are tilted in opposite directions. The average $|\delta
B_{h}|$ per timestep that is injected by the driving is only half
the value given by equation (\ref{eq:dBh}), because one in four
steps cause the stresses to decrease (Fig.~\ref{distribution}).
Noting that $B_{c} = B_{v} \tan \theta_{c}$, we find that the
waiting time is given by

\be T_{w} \approx \frac{B_{c}/2}{\delta B_{h}/2} \approx \frac{L
\tan \theta_{c}}{d} \label{tw_eq4} \ee

\noindent in units of the timestep duration (1000 s).  In
Figure~\ref{twait} we show log-log plots of $T_{w}$ versus the
parameters.  The scalings are: $L^{0.88}$ and $(\tan
\theta_{c})^{0.85}$.  This is generally consistent with equation
(\ref{tw_eq4}), and we suggest that small deviations are due to the
subjective nature of identifying the start of the main phase. Notice
that $T_{w}$ is expected to be independent of $B_{v}$, which we have
confirmed to be true from our simulations.  We emphasize again that
the waiting time is different from the rise duration. The rise
duration is more closely related to the spread of the driver
distribution (e.g., the Gaussian width in Fig.~\ref{distribution}b)
than it is to the time required for the average strand to reach
critical.

The duration of the decay phase depends on both the form of the
final driver distribution (Fig.~\ref{distribution}b) and the
timescale for transitioning from the initial to final distributions.
We leave the details for a later study.


\section{Discussion}
\label{discussion}


We develop a pseudo-1D model based on a cellular automaton (CA) that
reproduces the main properties of the evolution of coronal loops
observed in soft X-rays. As we describe in Section~\ref{model}, the
model is based on the idea that coronal heating is due to the sudden
reconnection of tangled magnetic strands (Parker 1988). Using
typical solar ranges for the model parameters we qualitatively
reproduce the observed intensity amplitude, fluctuations, and
evolution timescales of soft X-ray loops studied in L\'opez Fuentes
\etal (2007). The synthetic loops obtained with the CA+EBTEL model
combination also reproduce the measured physical properties of the
observed loop plasma such as density and temperature.


Although our model is related to usual models of self-organized
criticality (SOC, see e.g. Charbonneau \etal 2001), it differs in
some important aspects. In most SOC models the rules for the
critical condition and the redistribution of the field among
neighboring strands are based on mean values of the field strength.
The strength of each strand is compared with the mean value of its
neighbors, and if the critical condition is fulfilled the field is
redistributed equally among those neighbors. One may classify this
type of model \textit{isotropic}. As we describe in
Section~\ref{model}, our CA model is \textit{anisotropic} in this
respect. The other important difference is that SOC models use null
boundary conditions. The mesh points (strands) at the boundary are
set to have zero magnetic field (what we call $B_{h}$ in
Section~\ref{model}) at all times. Here, we use instead periodic
boundary conditions. It may be partly due to these differences that
the nanoflares produced with our model do not follow power law
distributions, which are characteristic of SOC models (for an
anisotropic SOC model associated with power law distributions, see
e.g. Morales \& Charbonneau 2008). We must stress that our main goal
here is not to obtain such distributions, but instead to explain the
evolution of loops in the frame of the nanoflare model. It is worth
pointing out, though, that our CA model reaches a steady critical
state in much the same way as SOC models do (see Figure~\ref{example}).


As we mention in Section~\ref{intro}, in recent years there has been
a debate over the issue of whether coronal loops are isothermal or
multithermal. The argument goes that if loops are isothermal they
are likely to be monolithic structures, while multithermality
implies that loops are composed of different unresolved strands
evolving independently. There seems to be evidence in favor of both
kinds of observations (see e.g., Schmelz \etal 2009), so a consensus
is forming that both scenarios actually occur. Let us analyze this
possibility in terms of the model presented here. For instance, if
the reconnection mechanism is not very efficient (e.g., a small $f$
factor, see Section~\ref{model}) and all nanoflares are
approximately the same size, then the heating injection will be
difficult to distinguish from a single source of constant heating.
In that case, the strand will keep more or less the same temperature
along the main phase of the evolution. If all the strands that
comprise the loop have similar physical conditions (it is expected
that they all have approximately the same field strength, for
example), then the loop will be nearly isothermal. These loops can
be easily taken as examples of quasi-static evolution. On the other
hand, if nanoflares occur less frequently, so the waiting time
between consecutive nanoflares is not negligible, the strands have
time to cool to lower temperatures between one nanoflare and the
next. Since nanoflares occur randomly, at any given time during the
evolution of the loop there will be strands simultaneously having
different temperatures. That would correspond to a multithermal loop
observation.

There is yet a third possibility.  If the nanoflares occur only once
in each strand, without repeating, and if they all occur at
approximately the same time, then the loop will appear roughly
isothermal at any given moment.  The temperature will decrease on a
cooling timescale, however, and the loop will be short lived.  We
call this scenario a short duration nanoflare ``storm" (Klimchuk
2009). Many warm EUV loops can be explained in this way.  Hot soft
X-ray loops tend to live much longer than a cooling time, on the
other hand (L\'opez Fuentes \etal 2007). If they are indeed
isothermal, then they can only be explained by rapidly repeating
nanoflares or by truly steady heating.


The three alternative scenarios described above relate to the
question weather the same loop is observable by instruments that are
sensitive to different temperatures.  In the case of nanoflares that
repeat rapidly in each strand of the loop bundle, the loop will be
visible only over a narrow range of temperatures.  It will be seen
either in hot soft X-ray emission or in warm EUV emission, but not
both. In the case of nanoflares that repeat with a time delay longer
than the cooling time, the loop will be simultaneously visible in both
emissions.  In the final case of a short duration nanoflare storm, the
loop will be visible first in soft X-rays and then in the EUV.

Observational evidence has been presented for all three
possibilities.  The question of which is most prevalent is not yet
answered.  Loops produced by short duration storms have been clearly
identified and can be found at many locations within active regions
(e.g., Ugarte-Urra \etal 2006, 2009).  On the other hand, soft X-ray
loops are most pronounced in the central cores of active regions,
where EUV loops are less common (Schmieder \etal 2004).  There are
also reports of soft X-ray and EUV loops that are almost but not
exactly co-spatial (Nitta 2000, Nagata \etal 2003). We attempted to study
the warm emission associated with the SXI loops studied in L\'opez
Fuentes \etal (2007).  Unfortunately, TRACE data are not available,
so we had to rely on lower cadence observations from SOHO/EIT. Since
only four images per day were taken in each channel, we could only
compare the hot and warm emission at a few times during the loop
evolution. We found that some of the SXI loops had an EIT
counterpart but others did not. Clearly, more systematic
observations of loops in different wavelengths are needed.


It is possible that there are two well differentiated populations of
AR loops. Hotter and longer-lived loops are preferentially located
at the core of ARs (Antiochos \etal 2003, Ugarte-Urra \etal 2009),
while the periphery is dominated by longer, shorter-lived,
cooler loops (Del Zanna \& Mason 2003). If this is indeed the case
in general, then the different locations and evolutions of these
loop classes may be related to different physical properties at
their origin. The model we present in this paper applies
specifically to hot loops. Whether it can explain warm loops at the
periphery of ARs is unclear. We can nonetheless speculate on the
difference in these two regions. The strength of the coronal
magnetic field is generally greater in the core of an AR and this
will tend to produce stronger heating. Our model also depends
fundamentally on the properties of footpoint shuffling. It may be
that these properties vary systematically across the active region.
This is an important question that we hope will be answered with the
latest magnetogram observations from SOT/Hinode. These observations
will hopefully also confirm our proposal that the slow decay
phase of hot loops is associated with the dispersal of magnetic
footpoints to the point where magnetic field tangling is no longer
efficient.


In Section~\ref{results} we found a series of scaling laws relating
observed properties of the loop evolution to different physical
parameters of the model. Some of these parameters, such as the loop
length, can actually be obtained directly from the observations. The
strength and tilt of the magnetic field at the coronal base can be
inferred by combining photospheric magnetogram observations with
models of the coronal structure (see e.g., L\'opez Fuentes \etal
2006, Mandrini \etal 2000). Therefore, these scaling laws are
predictions of the model that can be tested observationally.


From the discussion in the previous paragraphs we conclude that
among future important investigations is the systematic study of
loop observations in different wavelengths and the analysis of
possible scalings of the evolution parameters (timescales, mean
intensities, fluctuations) with the loop physical properties such as
length and magnetic field. From the theoretical side, the model
presented here is still very simple, and more sophisticated versions
should be developed, such as 2D models that better account for the
vector nature of the magnetic field and the fact that each magnetic
strand is wrapped around multiple other strands. We have already
started work in this direction.


\acknowledgements The authors wish to thank Paul Charbonneau for enriching
discussions and the anonymous referee for useful comments and suggestions.
This work was supported by the NASA Supporting Research and Technology
and Living With a Star Programs.


\clearpage


\begin{table}
\caption{Numerical values used for the parameters of the model.}
\vspace{0.5cm}
\label{table1}
$\begin{array}{lccccccc}
\hline
B_{v} (Gauss)   & 20  & \bf{40} & 60 & 80 & 100 & 120 \\
L (Mm)          & 20  & 40 & 60 & 80 & \bf{100} & 120 \\
\tan \theta_{c} & 0.1 & \bf{0.2} & 0.4 & 0.6 & 0.8 & 1.0 \\
f               & 0.1 & \bf{0.2} & 0.4 & 0.6 & 0.8 & 1.0 \\
N_{m}           & 20  & 50 & \bf{100} & 200 & 500 & 800 & 1000 \\
\tau            & 100 & 500 & \bf{1000} & 2000 & 3500 & 5000 \\
\hline
\end{array}$
\end{table}

\begin{table}
\caption{Compared properties of the observed and modeled lightcurves shown if Figure~\ref{lightcurves}.}
\vspace{0.5cm}
\label{table2}
$\begin{array}{lcccc}
\hline
   & \multicolumn{2}{c}{$Loop 1$}   & \multicolumn{2}{c}{$Loop 4$}     \\
\hline
                                & $Obs.$ & $Model$ & $Obs.$ & $Model$  \\
\hline
$Rise phase duration (hr)$          & 3.35  & 3.65 & 3.74  & 3.16    \\
$Rise phase timescale (hr)$         & 1.97  & 1.88 & 2.17  & 1.75    \\
$Main phase intensity (DN/pix.sec)$ & 6.82  & 7.59 & 2.99  & 2.82    \\
$Main phase RMS$                    & 0.10  & 0.10 & 0.13  & 0.09    \\
$Decay phase duration (hr)$         & 2.98  & 3.99 & 4.51  & 4.19    \\
$Decay phase timescale (hr)$        & -1.51 & -2.20 & -2.47 & -2.48  \\
\hline
\end{array}$
\end{table}


\clearpage

\clearpage
\begin{figure*}
\centering
\hspace{0.cm}
\includegraphics[bb= 0 231 1064 842,width=17cm]{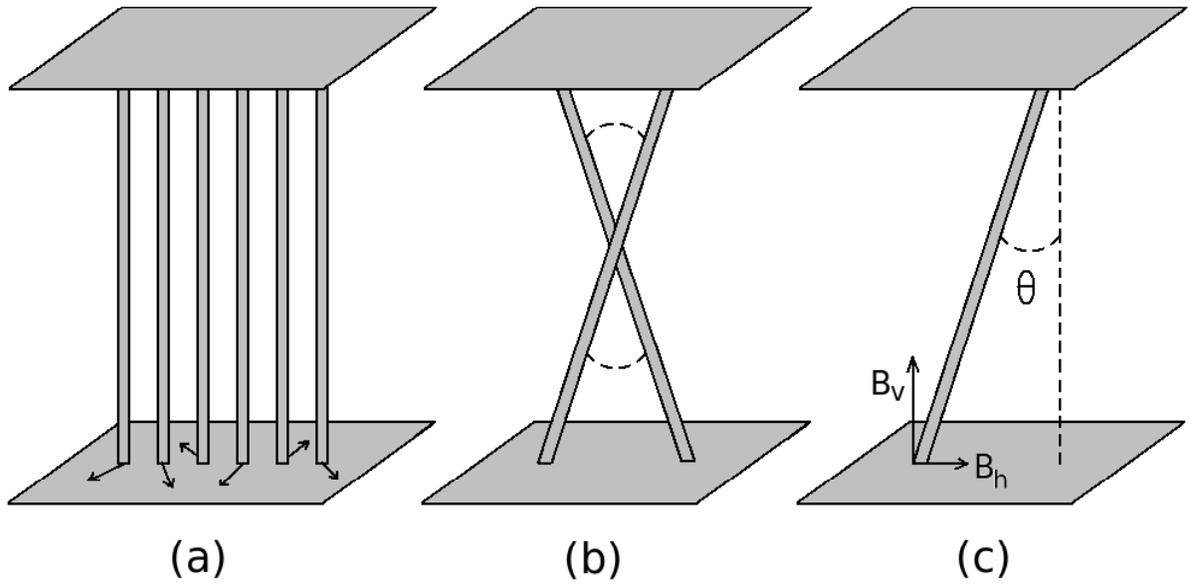}
       \caption{Schematic description of the CA model. (a) Initially, all strands
       are
parallel and there is no magnetic stress between neighbors. The
horizontal planes represent two distant portions of the photosphere
connected by the strands. (b) Photospheric motions displace the
strand footpoints increasing the mutual inclinations and producing
magnetic stress. (c) We associate the inclination of the strands
($\theta$) with the appearence of a horizontal component of the
magnetic field (see Section~\ref{description}).}
         \label{drawing}
\end{figure*}

\clearpage
\begin{figure*}
  \centering
\hspace{0cm}
\includegraphics[bb= 130 30 465 740,width=9cm]{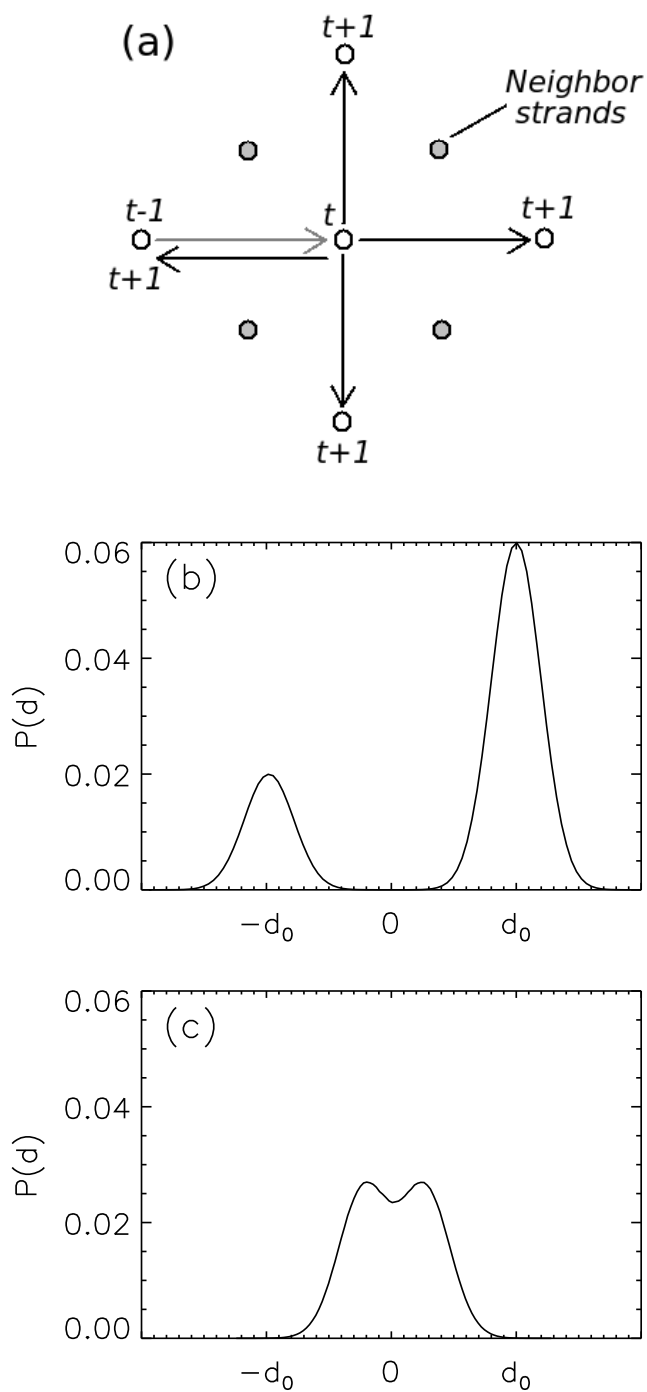}
      \caption{Panel (a): the gray arrow indicates the footpoint
      displacement
from time step $(t-1)$ to time step $t$ and the black arrows
indicate the possible further evolution from timestep $t$ to
timestep $(t+1)$, which may increase (3/4 probability) or cancel
(1/4 probability) the displacement of the previous step. The driver
distribution of panel (b) is consistent with these probabilities. To
simulate the footpoint dispersion, the distribution of panel (b)
slowly evolves to the distribution of panel (c) (see
Section~\ref{implementation})}.
         \label{distribution}
\end{figure*}

\clearpage
\begin{figure*}
  \centering
\hspace{0cm}
\includegraphics[bb= 54 360 535 1068,width=14cm]{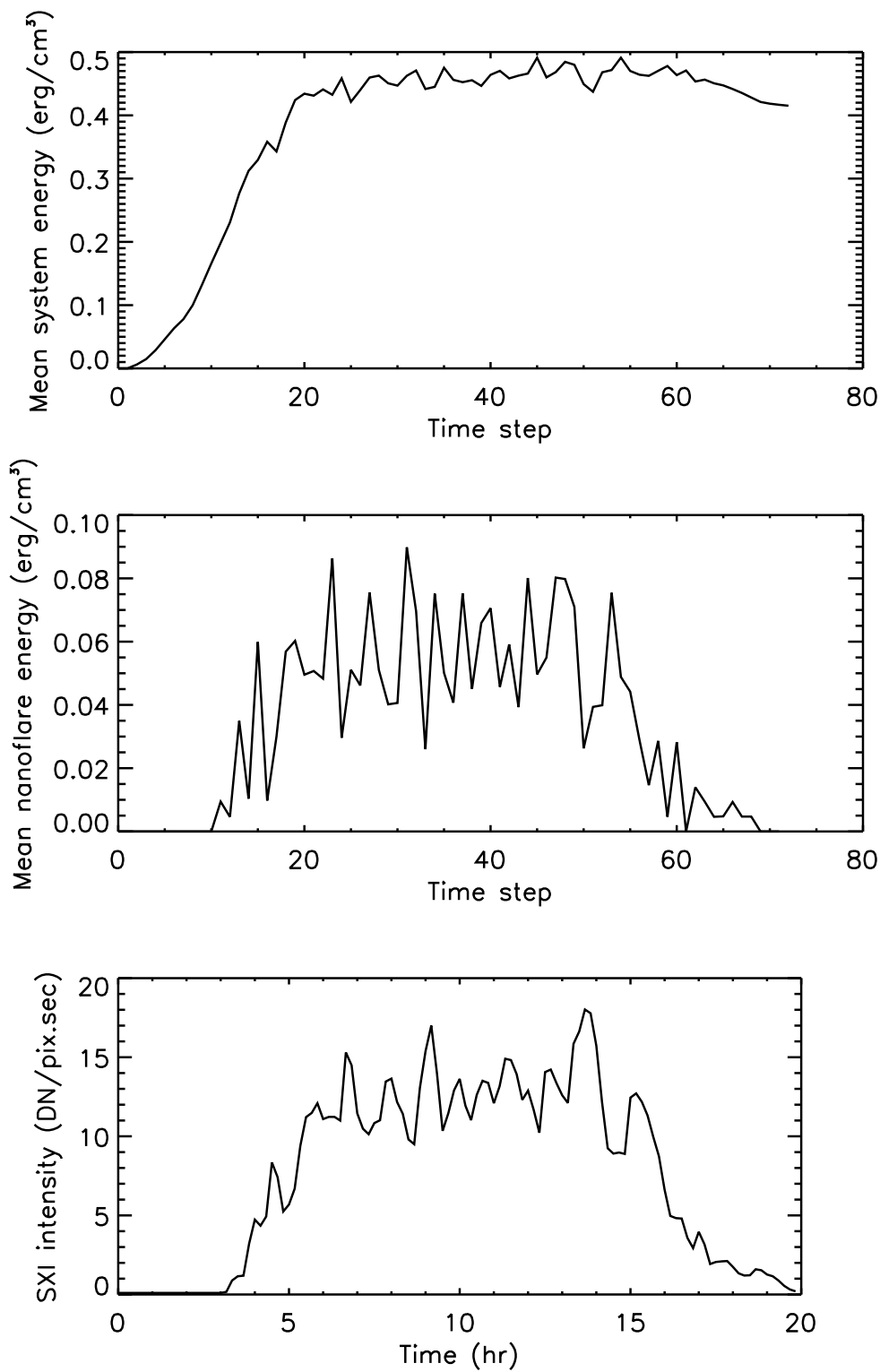}
      \caption{Example of the CA model evolution. Upper panel: mean magnetic
      energy
(per strand) versus time step number. Middle panel: idem for the
mean nanoflare energy (per strand, see
Section~\ref{implementation}). Lower panel: light curve obtained
from the model of the upper panels (see Section~\ref{plasma}). The 3
panels have a precise vertical alignment, so that 20 hr corresponds
to time step 72.}
         \label{example}
\end{figure*}

\clearpage
\begin{figure*}
\centering
\hspace{0.cm}
\includegraphics[bb= 14 14 740 653,width=16cm]{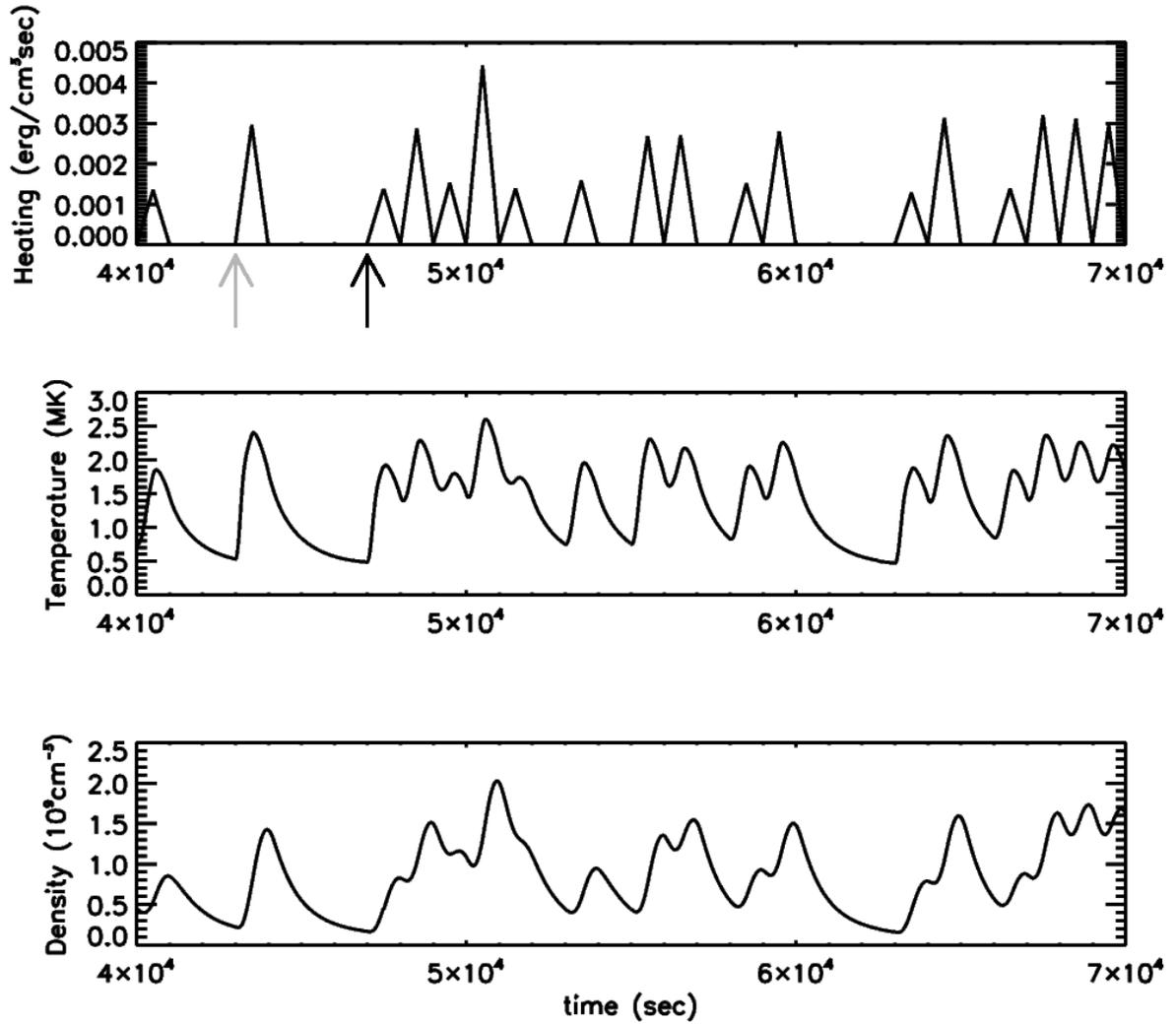}
       \caption{Plasma response to a series of nanoflares modeled with the EBTEL code.
The gray arrow indicates the start time of a isolated nanoflare,
while the black arrow corresponds to the start of a ``train'' of
events occurring in a rapid succession (see Section~\ref{plasma}).}
         \label{ebtel}
\end{figure*}

\clearpage
\begin{figure*}
\centering
\hspace{0.cm}
\includegraphics[bb= 60 85 550 770,width=12cm]{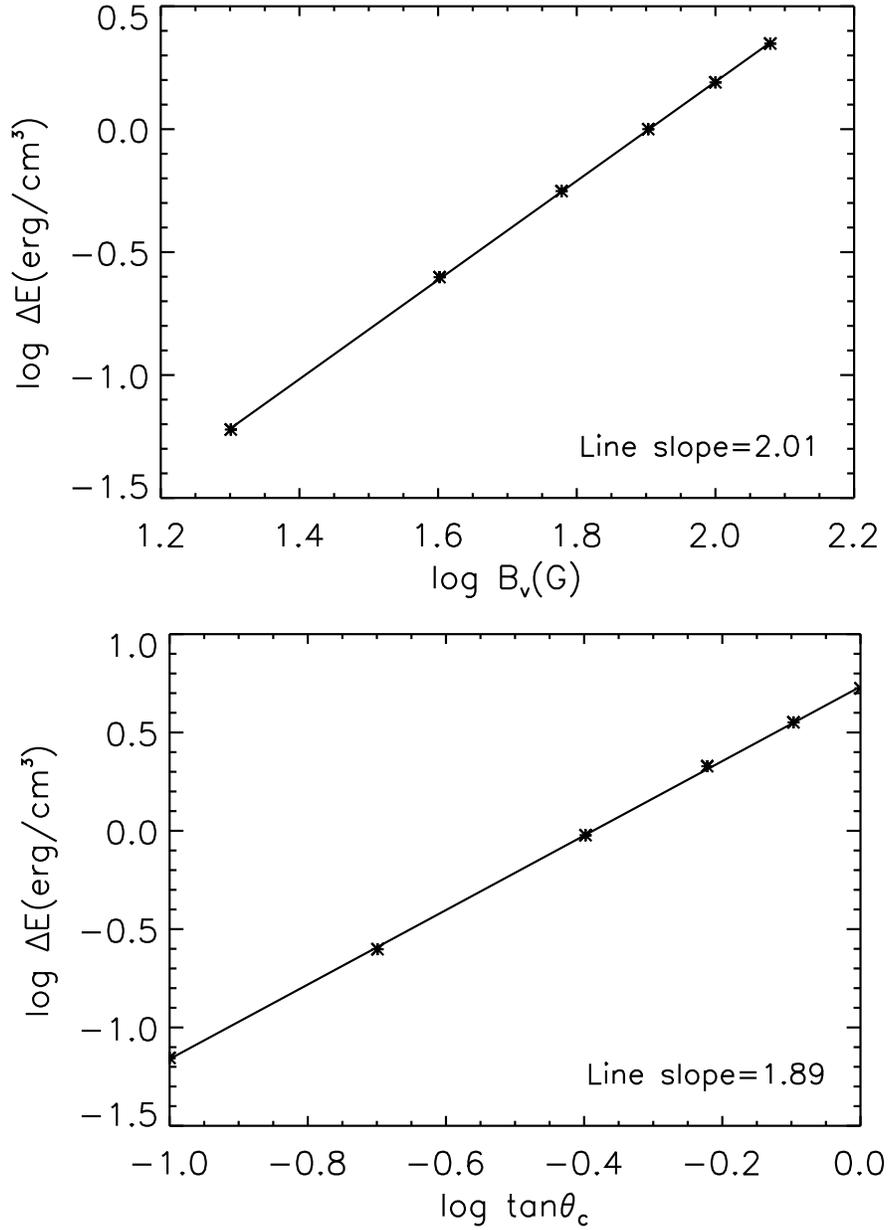}
       \caption{Log-log plots of the mean reconnection event energy, $\Delta E$, versus relevant parameters of the model (see Section~\ref{results}).}
         \label{energy}
\end{figure*}

\clearpage
\begin{figure*}
\centering
\hspace{0.cm}
\includegraphics[bb= 130 70 470 780,width=9cm]{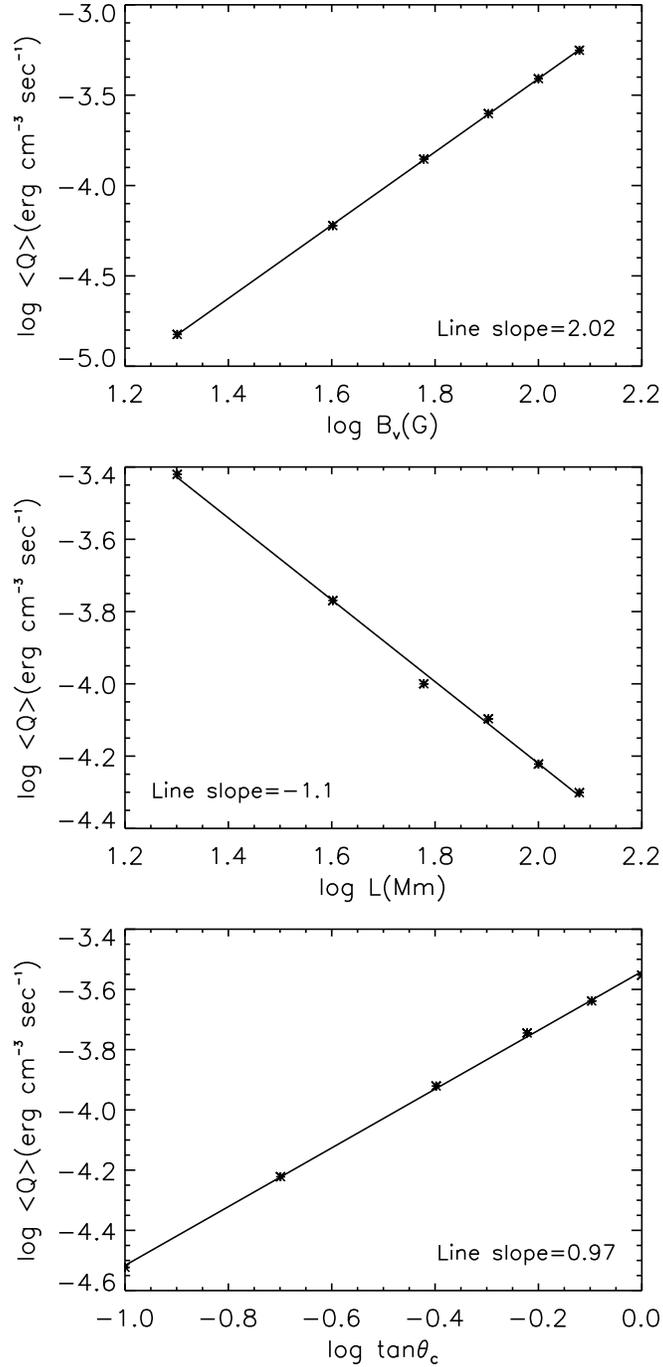}
       \caption{Log-log plots of the mean nanoflare heating rate, $\left< Q \right>$, versus parameters of the model (see Section~\ref{results}).}
         \label{nanoflares}
\end{figure*}

\clearpage
\begin{figure*}
\centering
\hspace{0.cm}
\includegraphics[bb= 130 60 440 780,width=9cm]{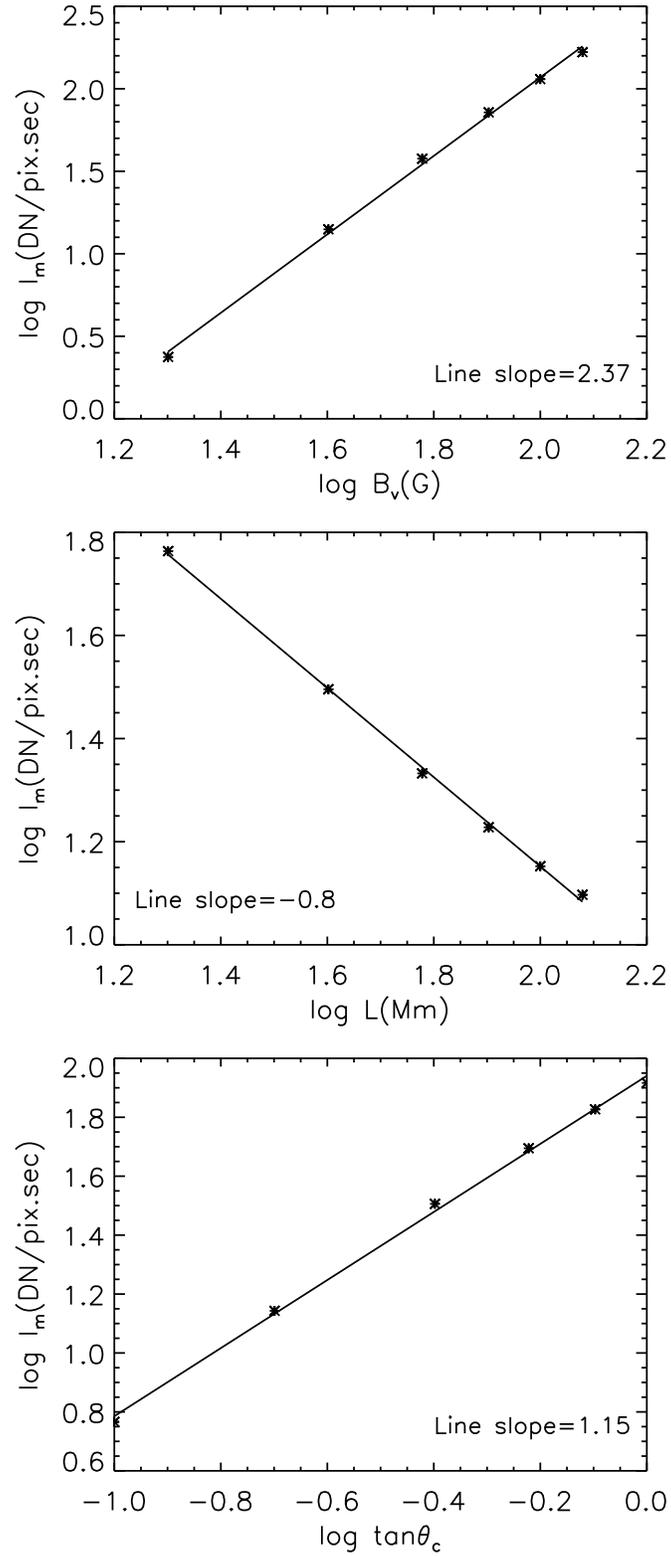}
       \caption{Log-log plots of the main phase mean intensity, $I_{m}$, versus relevant parameters of the model (see Section~\ref{results}).}
         \label{main}
\end{figure*}

\clearpage
\begin{figure*}
\centering
\hspace{0.cm}
\includegraphics[bb= 100 50 450 790,width=9cm]{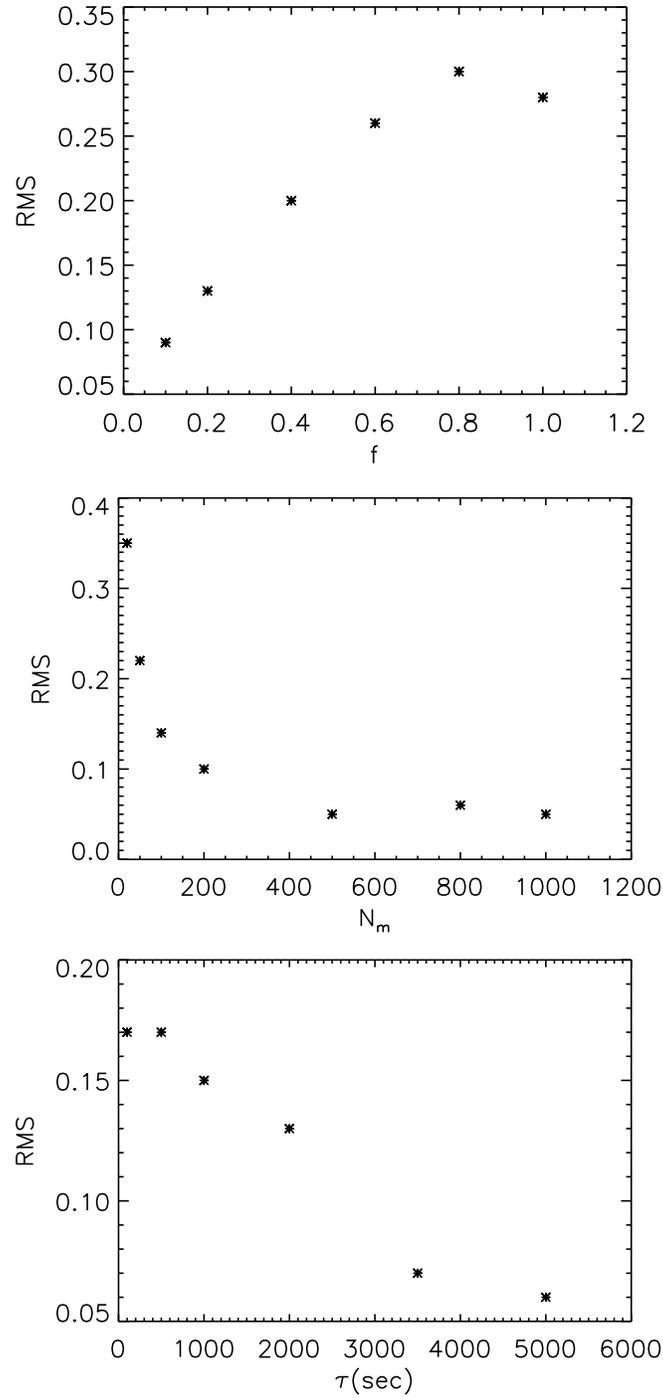}
       \caption{Scatter plots of the root mean square variation of the main phase intensity, $RMS$, versus relevant parameters of the model.}
         \label{rms}
\end{figure*}

\clearpage
\begin{figure*}
\centering
\hspace{0.cm}
\includegraphics[bb= 90 90 530 750,width=9cm]{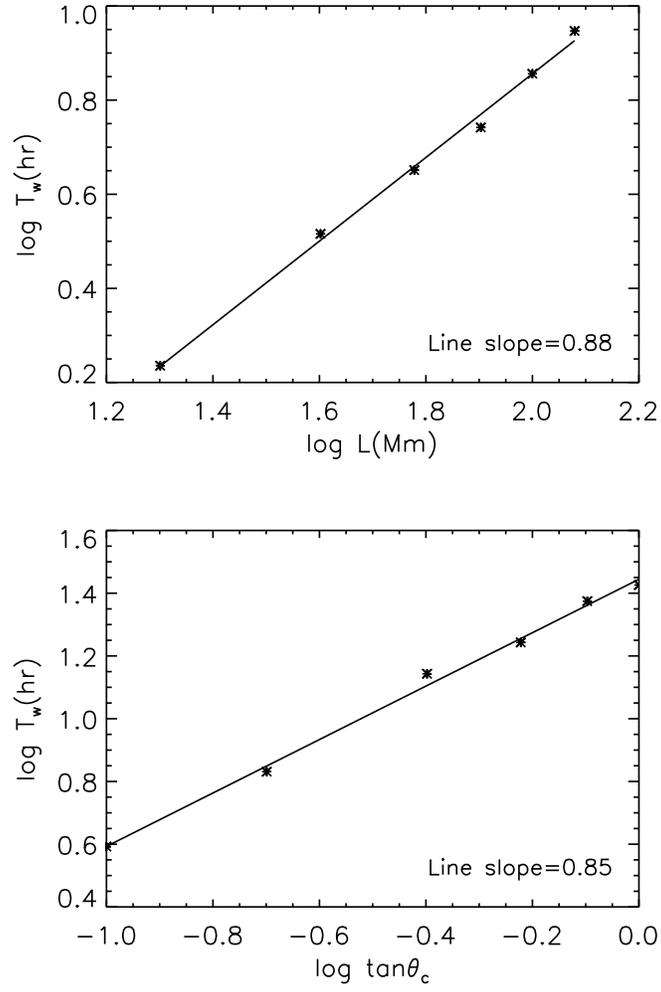}
       \caption{Log-log plots of the waiting time, $T_{w}$ (see Section~\ref{results}), versus relevant parameters of the model.}
         \label{twait}
\end{figure*}

\end{document}